\documentclass[]{aa}
\usepackage{graphicx}
\usepackage[intlimits,sumlimits]{amsmath}
\usepackage{deluxetable}
\usepackage{times,epsfig} 
\usepackage{natbib}
\bibpunct{(}{)}{;}{a}{}{,} 
\usepackage{amsmath}
\usepackage[english]{babel}

\newcommand{\beqa}{\begin{eqnarray}} 
\newcommand{\eeqa}{\end{eqnarray}}
\newcommand{\Abst}[1]{\;#1}
\newcommand{\bsub}{\begin{subequations}}
\newcommand{\esub}{\end{subequations}}
\newcommand{\beal}{\begin{align}}
\newcommand{\ealn}{\end{align}}
\newcommand{\Nif}{$\rm ^{56}Ni$} 
\newcommand{\Cif}{$\rm ^{56}Co$}
\newcommand{\Fif}{$\rm ^{56}Fe$}
\newcommand{\ksm}{${\rm km~s^{-1}~Mpc^{-1}}$}
\newcommand{\s}{M$_{\sun}$}
\begin{document}

\title{Constraints on the Progenitor Systems of Type~Ia Supernovae}
\titlerunning{Bolometric properties of SNe~Ia}

\author{M.~Stritzinger\inst{1}
\thanks{Current address and E-mail: Dark Cosmology Centre, Niels Bohr 
Institute, University of Copenhagen, Juliane Maries Vej 30, DK-2100 
Copenhagen \O, Denmark (max@astro.ku.dk)}
	\and B.~Leibundgut\inst{2} 
   	\and S.~Walch\inst{3}
	\and G. Contardo\inst{1}
}

\institute{Max-Planck-Institut f\"ur Astrophysik, Karl-Schwarzschild-Str. 1,
85741 Garching bei M\"unchen, Germany \\
\email{stritzin@mpa-garching.mpg.de}
\and European Southern Observatory, Karl-Schwarzschild-Str. 2, 85748
Garching bei M\"unchen, Germany \\
\email{bleibundgut@eso.org}
\and Universit\"atssternwarte M\"unchen, Scheiner-Str. 1, 81679 M\"unchen, 
Germany \\
\email{swalch@usm.uni-muenchen.de}}


\offprints{M. Stritzinger} 

\date{Received; Accepted }

\abstract {UVOIR bolometric light curves provide a valuable
insight into the nature of type Ia supernovae. 
We present an analysis of sixteen well-observed SNe~Ia. 
Constraints are placed on several global parameters concerning the
progenitor system, explosion mechanism and subsequent radiation transport. 
By fitting a radioactive decay energy deposition function to the 
quasi-exponential phase (50 to 100 days after maximum light), it is
found that the ejected mass varies by at least a factor of two. 
This result suggests that a sub Chandrasekhar mass white dwarf could
be the progenitor system of some SNe~Ia.
We find that the range in the amount of synthesized \Nif~indicates 
a significant variation in the burning mechanism. In order to explain a
factor of ten range in the observed bolometric luminosity, more
detailed modeling of the explosion mechanism is required. 
\keywords{supernovae: general}}
\titlerunning{Bolometric properties of SNe~Ia}
\authorrunning{M. Stritzinger et al.}

\maketitle
\section{Introduction}
Type Ia supernovae (hereafter SNe~Ia) have become
an exceptional tool in modern cosmology. Due to their large luminosity, 
they are used to place constraints on cosmological parameters,
and as of yet, provide the only {\it direct} evidence for the existence 
of dark energy \citep{riess98,perlmutter99,leibundgut01}. 
Despite the insight SNe~Ia have given us about the universe,
several key issues related to the nature of their
progenitor system(s) and the physics of the explosion mechanism(s) have 
remained unsolved \citep[for reviews see][]{hillebrandt00,livio00}.

Today it is commonly believed that SNe~Ia emerge from the 
thermonuclear incineration of a carbon oxygen (C-O) white dwarf 
exploding near or at the Chandrasekhar mass 
\citep{nomoto84,woosley86,hillebrandt00}. 
The energy released from burning to nuclear statistical equilibrium
(NSE) completely destroys the white dwarf. The
optical/IR light curves are powered by the Comptonization of 
$\gamma$ rays produced from the radioactive decay chain  
$^{56}$Ni~$\rightarrow$~$^{56}$Co $\rightarrow$~$^{56}$Fe
\citep{pankey62,colgate69}. Within this paradigm the C-O white dwarf 
accretes matter from an evolved massive star either via Roche lobe overflow 
or through stellar winds. With the detection of H$\alpha$
in the well-observed SN~2002ic \citep{hamuy03,nomoto04,kotak04}
the single degenerate model has become favored over the double degenerate 
model. 

It is clear that SNe~Ia are not a homogeneous group
of stellar explosions rather, they display a range in luminosity of
a factor of ten or more \citep{suntzeff96,contardo00,suntzeff03,stritzinger05}. 
In fact not one self-consistent explosion model has yet been presented 
that can successfully account for this observed range in luminosity. 
This lack of understanding of the 
progenitor systems is unsettling, and must be addressed, if we are to 
have confidence in the cosmological results provided by SNe~Ia.

Fortunately, during the past decade, a number of observing campaigns 
\citep[see][ for a list]{leibundgut00} have obtained
excellent data sets for a large number of SNe~Ia. 
With these data sets we are currently in a position 
to conduct a systematic investigation of their photometric 
and spectroscopic properties. 
The motivation of this work is 
to place further constraints on both the progenitor systems 
and explosion mechanisms with these observations. 
In particular, we construct UVOIR bolometric
light curves from broad-band photometry of a number of SNe~Ia. 
With the UVOIR light curve we determine several global parameters, 
e.g. total ejected mass, \Nif~mass and the $\gamma$-ray escape fraction. 
Investigating the range of such parameters can provide insights
into the nature of SNe~Ia, and help differentiate between the
various paths of stellar evolution that the progenitors follow,
as well as the manner in which thermonuclear combustion occurs.

The structure of this paper is as follows. Sect.~2 
discusses the observational data used and the UVOIR light curves.
Sect.~3 describes our analysis of the UVOIR light curves. 
Sect. 4  contains the results of this analysis and we conclude 
in Sect. 5.
\section{Observational data and UVOIR light curves}

($U$)$BVRI$-band observations for sixteen
SN~Ia have been procured from a variety of sources. Table~1 lists all events 
considered in this study along with references to the sources 
of the data, as well as a number of important parameters used 
and determined in this work. 
All events listed in Table~1 have excellent optical 
photometry that extends from pre-maximum out to $\sim100$ days past
maximum light. (See \citealt{stritzinger05} for more details concerning 
the method and sample.)

To determine the amount of \Nif~produced in the explosion from the 
observed UVOIR flux, two 
parameters are required. They include an estimate for the 
total extinction and the distance to the host galaxy. 
Values listed for Galactic reddening 
are those given by the COBE dust maps of \citet{schlegel98}.
Host galaxy reddenings were selected from \citet{phillips99} for those
SNe~Ia that coincide with our sample. For more recent events we adopted
host galaxy extinctions from the literature, with preference to those 
calculated by the Phillips method. The distances chosen depended mainly 
on what was available in the literature. For SNe~Ia without a direct distance
measurement, in this case, a Cepheid distance measurement or a
surface brightness fluctuation (SBF) distance we used a 
Heliocentric velocity obtained from NED and 
converted this to the CMB reference frame.\footnote{Through out this work
we have adopted H$_{\circ}~=~72$~\ksm.} For each CMB distance
we have assumed an uncertainty of 300~km~s$^{-1}$.

With the advent of the {\em Hubble Space Telescope} (HST) there has 
been a substantial effort from two groups
(namely the HST Key Project (HKP) and the Saha, Tammann and 
Sandage (STS) group) to obtain accurate Cepheid distance measurements to 
galaxies that have hosted SNe~Ia. 
Typically these two independent groups determine different distances for
any one galaxy, even though they use the same data and similar data reduction 
software. 
These differences are a reflection of assumptions made in 
their analysis. The main factor that contributes to these discrepancies is the 
exact $P$-$L$ relation used, and several
other subtle nuances exacerbate the problem. These 
include: (1) the criteria 
adopted to select the Cepheids used to determine the distance, 
(2) if and how metallicity corrections are applied, and (3) 
anomalies related to the camera(s) on HST \citep[see][]{parodi00,gibson00,riess05}. 

In short, the HKP obtains a {\em short} distance scale that
leads to a value of H$_{\circ}~\sim70$~\ksm while the STS group 
determines a {\em long} distance scale that gives 
H$_{\circ} \sim60$ \ksm. 
\citet{riess05} have reviewed this issue, in order to reconcile the distances
obtained by the two groups to galaxies that have hosted a SN~Ia. 

Four of the SNe~Ia used in our study have a direct Cepheid distance 
to their host galaxy and one has a Cepheid distance to its galaxy cluster.
Whether the distance scale is long or short will depend on which data
(i.e. from which of these two groups) we use.
This in turn will lead to either
an under or over estimate of the distance. 
This uncertainty in the distance will affect the \Nif~mass 
we determine and hence the ejected mass.

To illustrate the effect of this on our results, we consider the 
galaxy NGC 3982. This galaxy has three independent Cepheid distance 
measurements,
one from each of the two teams just mentioned and one from \citet{riess05}
who have used a new calibration of the $P$-$L$ relationship and an
elegant metallicity correction.

The STS group has published a distance modulus of $\mu = 31.72\pm0.14$
\citep{saha01b}, which is based on the Cepheid $P$-$L$ relation 
published by \citet{madore91}. Using a $P$-$L$ relation based 
on $\sim$650 Cepheids located in the Large Magellanic Cloud that
were observed by the Optical Gravitational Lensing Experiment (OGLE)
\citep{udalski99} the HKP team concluded that 
$\mu = 31.56\pm0.08$ \citep{stetson01}. More recently \citet{riess05}
used the Advanced Camera for Surveys on the HST to obtain 
a Cepheid distance. They used the  
$P$-$L$ relation presented by \citet{tammann02} and \citet{thim03},
based on only those Cepheids in the OGLE data set 
that have periods longer than 10 days. With this new $P$-$L$ relation, 
and after applying a metallicity correction, \citeauthor{riess05} found
$\mu = 31.66\pm0.09$.

With the HKP distance modulus we obtain a \Nif~mass that is 9\%
less compared to the STS distance modulus, and 
4\% less with the \citeauthor{riess05} distance modulus.
In turn the ejected mass derived with the HKP distance modulus is
22\% greater compared to the STS distance modulus, 
and 13\% greater compared to the \citeauthor{riess05} distance modulus.

We have used a SBF distance for three SNe~Ia in this sample. 
The zero-point for this method
is based on an empirical relation derived from Cepheids, which   
can lead to a systematic under- or over- estimation of the distance,
depending on whose Cepheids are used for the calibration. The three
SBF distances we have used are calibrated with the HKP Cepheid distances of
\citet{ferrarese00}. As pointed out by \citet{tonry01} the zero-point
for the SBF distance scale is still being improved. With the application
of the Cepheids determined with the complete set of OGLE data the 
SBF distances used here would be $\sim$0.1 mag fainter. With this
correction our derived \Nif~masses would increase by 
$\sim$8\% and the ejected masses would decrease by 
$\sim$16\%.

Finally, in Table~1 we list two parameters calculated in this study:
(1) an estimate of the amount of \Nif~produced from burning to NSE 
(see Sect.~\ref{nimass1}), and (2) an estimate of the time when the ejecta 
makes the transition from being optically thick to optically thin 
(see Sect.~\ref{ejectedmass1}).

To construct the UVOIR light curves we used the method 
already employed by \citet{vacca96,vacca97}, \citet{contardo00} 
and \citet{stritzinger05a}. We refer the reader to these articles for
a more detailed description of this empirical fitting method, which
we briefly summarize here.

Each filtered light curve is fitted with a ten-parameter function.
This function consists of a Gaussian for the peak phase, a linear
decline for the late-time decay (i.e. $^{56}$Co $\rightarrow$ $^{56}$Fe), 
an exponentially rising function to fit the initial rise 
to maximum, and a second Gaussian for the inflection or secondary maximum 
that is observed in the $VRI$-band light curves. 

SNe~Ia without a $U$-band light curve, a correction
was added, as described by \citet{contardo00}. \citeauthor{contardo00} used 
a correction based on SN~1994D \citep{richmond95,meikle96,smith00}, however, 
this event had an unusual blue color at maximum. For this reason, corrections 
based on SN~1994D 
tend to overestimate the fraction of flux associated with the $U$-band
photometry \citep{stritzinger05}. Instead we employed a correction derived 
from SN~1992A \citep{suntzeff96}. 
Although there are many well observed events, SN~1992A is one of the 
only {\it normal} SN~Ia that does not suffer extinction due to 
host galaxy reddening.

Owing to a lack of data we did not include 
the fraction of flux associated with wavelengths above
10,000~\AA. However, at maximum light the $JHK$-bands contribute no 
more than $\sim5\%$~\citep{suntzeff96} to the total bolometric flux.
Around 60 days after maximum light, when the bolometric light curve 
follows nearly a linear decline, the infrared contribution rises to 
no more than $\sim10\%$~of the total bolometric flux \citep{contardo01}.

To produce the UVOIR light curve, each fitted 
light curve is converted to flux (ergs s$^{-1}$). 
Next a reddening correction is 
applied, and then each filtered light curve is summed to obtain
the total flux. Note that we do not normalize the flux to any 
decline rate relation (e.g. $\Delta$m$_{15}$ \citep{phillips99}, MLCS
\citep{riess96} or stretch (\citealt{perlmutter97,nobili05}).

\section{Determining global parameters of SNe~Ia}
In this section we describe the manner in which UVOIR
light curves are used to determine the parameters of interest here.

\subsection{\Nif~mass}\label{nimass1}

At maximum light the peak luminosity of a SN~Ia is 
related, to first order, to the amount of \Nif~produced during the 
explosion. The amount of \Nif~synthesized from burning
to NSE is itself thought to be largely dependent on the 
explosion mechanism. With the \Nif~mass we are
directly probing the most sensitive part of the explosion and
can use observations to place constraints on the explosion 
mechanism.

\citet{suntzeff96} showed
that at maximum light $\sim$80\% or more of the total flux from a SN~Ia 
is emitted in the optical. Therefore with UVOIR light
curves, constructed from $UBVRI$ broad-band photometry, one easily
obtains a measure of the total flux and, through application of
Arnett's Rule, the \Nif~mass. Arnett's Rule states that at 
maximum light, the luminosity of a SN~Ia is equal
to the instantaneous energy deposition rate from the radioactive
decays within the expanding ejecta \citep{arnett82,arnett85}. 
To determine the \Nif~mass
we use the simple relation that gives for 1~M$_{\sun}$ of 
\Nif, a total luminosity at maximum light of
\beal
\label{eq:1.0}
{\rm L_{max}}~=~(2.0\pm0.3)~\times~10^{43}~\frac{{\rm M_{Ni}}}{{\rm M_{\sun}}}~~{\rm erg~s^{-1}}
\Abst{.}
\end{align}
The error in Eq.~(\ref{eq:1.0}) corresponds to a 3-day 
uncertainty in the adopted
bolometric rise time of 19 days \citep[see Sect. 4.1 of][ for more details]
{stritzinger05a}. As approximately 10\% of the 
total flux at maximum light is emitted outside of the optical, 
each \Nif~mass derived from Eq.~(1) has been increased
by a factor of 1.1. The dominant errors in the
deduced \Nif~mass are associated with the adopted distance to the host galaxy
and the total extinction \citep{contardo00}. 

\subsection{Total ejected mass}\label{ejectedmass1}
To place constraints on the ejected mass 
we perform a least-squares fit of a radioactive beta-decay energy (RDE)
deposition function to the post maximum phase UVOIR light curve. 
Prior works that discuss this method
include the pioneering investigations of \citet{colgate80a,colgate80b}, 
followed by the more sophisticated treatment presented by \citet{jeffery99}; 
(see also \citealt{cappellaro97}, \citealt{milne99} and 
\citealt{milne01} for similar methods and techniques.)
However, as of yet, no attempt has been made to apply such a method to UVOIR 
light curves derived from real observations.

An expression for the energy deposition of N$_{\rm {Ni0}}$ atoms
of \Nif~ in the optically thin limit (i.e. when $\tau \ll 1$) is
represented by 

\begin{multline}
\label{eq:2.0}
E_{{\rm dep}} =  E_{{\rm Ni}} + E_{{\rm Co~e^{+}}} + [1 - {\rm exp(-\tau)}]E_{{\rm Co~\gamma}} \\[2mm]= \lambda_{{\rm Ni}}{{\rm N_{Ni0}}}~{\rm exp(-\lambda_{{\rm Ni}}t)Q_{{\rm Ni~\gamma}}} \\[2mm]+ \lambda_{{\rm Co}}{\rm N_{Ni0}} {\frac{\lambda_{{\rm Ni}}}{\lambda_{{\rm Ni}}-\lambda_{{\rm Co}}}}[{\rm exp(-\lambda_{{\rm Co}}t)- exp(-\lambda_{{\rm Ni}}t)}]\\[2mm] \times \{Q_{\rm Co~e^{+}} + Q_{{\rm Co~\gamma}}[1 - {\rm exp(- \tau)}]   \}
\Abst{.}
\end{multline}

\noindent $\lambda_{{\rm Ni}}$ and $\lambda_{{\rm Co}}$ are the respective 
e-folding decay times of 8.8 and 111.3 days for $^{56}$Ni and $^{56}$Co.
$Q_{{\rm Ni~\gamma}}$ (1.75 MeV) is the energy released per 
$^{56}$Ni~$\rightarrow$~$^{56}$Co decay.
$Q_{{\rm Co~e^{+}}}$ (0.12 MeV) and 
$Q_{{\rm Co~\gamma}}$ (3.61 MeV) are the positron and $\gamma$-ray 
energies released per $^{56}$Co $\rightarrow$ $^{56}$Fe 
decay (for a detailed discussion of the 
radioactive properties of this decay chain see, e.g.  
\citealt{nadyozhin94}.)
Note that throughout this work we assume that all neutrinos produced 
from the~\Nif$\rightarrow$\Cif$\rightarrow$\Fif~decay chain escape the 
ejecta entirely and do not contribute to the observed UVOIR flux.

As Eq.~(\ref{eq:2.0})  is only applicable in the optically thin limit, 
when the thermalized photons can freely escape, 
it is safe to assume that at these epochs the majority 
of \Nif~has decayed to $^{56}$Co, and therefore the remaining 
amounts of \Nif~provides a negligible contribution to the energy 
deposition. At these 
epochs the UVOIR light curve appears to be nearly first order 
exponential, however it is more accurately described as `quasi-exponential'
\citep[see][$~$for a detailed discussion]{jeffery99}.
With the presence of only one radioactive species, the mean optical depth
$\tau$ has a simple t$^{-2}$ dependence:
\beal
\label{eq:3.0}
\tau = \frac{t^{2}_{\circ}}{t^{2}}
\Abst{.}
\end{align}
If we replace $\tau$ in Eq.~(\ref{eq:2.0}) with Eq.~(\ref{eq:3.0}), and 
then perform a least-squares fit of Eq.~(\ref{eq:2.0}) to the UVOIR 
light curve (between 50 and 100 days past maximum light 
when Eq.~(\ref{eq:2.0})  is valid), we can determine the 
`fiducial time' t$_{\circ}$. It is at this time that  
the ejecta becomes optically thin.

Following the discussion of \citet{jeffery99}, one finds that  
t$_{\circ}$ can be expressed as
\beal
\label{eq:4.0}
t_{\circ} = \left(\frac{M_{ej} \kappa q}{8 \pi}\right)^{\frac{1}{2}} \frac{1}{v_{e}}
\Abst{.}
\end{align}
The variable M$_{{\rm ej}}$ is the total ejected mass, $\kappa$ 
is the $\gamma$-ray
mean opacity, v$_{e}$ is the e-folding velocity of an exponential model's
density profile, and q is a general form factor that describes the 
distribution of \Nif~in the ejecta. 

During the optically thin phase for an all-metal ejecta ($\mu_{e} = 2$),
$\kappa$ is expected to be in the range  
0.025 to 0.033~cm$^{2}$ g$^{-1}$ \citep[see][$~$and references therein for a 
detailed discussion]{swartz95,jeffery99}. 
We adopted the value of 0.025~cm$^{2}$ g$^{-1}$ 
as our fiducial $\gamma$-ray mean opacity.

\citet{jeffery99} lists the model e-folding velocity 
of several successful 1-D explosion models consisting of
1.4 M$_{\sun}$ Chandrasekhar-size white dwarfs. These e-folding velocities 
are $\sim$2700~km~s$^{-1}$ for W7 \citep{nomoto84},
2750~km~s$^{-1}$ for DD4 \citep{woosley94a}, and 3000~km~s$^{-1}$
for M36 \citep{hoflich95}. In addition \citet{jeffery92} found
that the DD2 model of \citet{woosley91} has a e-folding velocity
of $\sim$3160 km s$^{-1}$. 

More recently \citet{ropke05} published two full-star 3-D explosion 
models of a 1.4 M$_{\sun}$ white dwarf; with different ignition conditions:
a centrally ignited configuration (c3\_4$\pi$) and a foamy multi-bubble flame
structure ($f$1). Using Eq.~(A10) of \citet{jeffery99} and parameters
given in Table~1 of \citet{ropke05}, we have calculated the
e-folding velocities for these two models. 
The e-folding velocities correspond to $\sim$1611 km~s$^{-1}$ for the 
c3\_4$\pi$ simulation and $\sim$1842 km~s$^{-1}$ for the $f$1 simulation. 
These values are substantially smaller than the previously cited 
1-D models and reflect the difference
between the density profiles generated by 1-D and 3-D simulations.
Note that all these models are based on  
the explosion of a 1.4~M$_{\sun}$ Chandrasekhar-size white dwarf.
In the calculations presented below we arbitrarily adopted 
3000~km~s$^{-1}$ as our `average' fiducial e-folding velocity. 

The parameter q is equal to one for high concentrations of 
\Nif~at the center of the ejecta, small for low concentrations within the 
center, and one-third for the case when the \Nif~is evenly 
distributed throughout the ejecta 
\citep[see][ for a detailed discussion]{jeffery99}.
There is mounting evidence that an appreciable amount of \Nif~is  
moderately mixed within the ejecta. However, it is likely that the amount 
of mixing may vary significantly from supernova to supernova.

An analysis of early-time spectra of SN~1991T
\citep{ruiz92,mazzali95} indicates the existence of an outer shell of \Nif.
In contrast, \citet{georgii02} presented observations of SN~1998bu obtained 
with COMPTEL. They concluded that their non-detection of $\gamma$ rays 
from the \Cif$\rightarrow$\Fif~decay chain indicates that there is  
no appreciable mixing of radioactive nuclides 
within the ejecta in the context of current models.
More recently \citet{stehle05} have presented ``abundance tomography" of
SN~2002bo. With their unique technique they determined 
that the vast majority of \Nif~was distributed between 
3000 to 11,000~km~s$^{-1}$ for this particular event. 
\citet{jeffery99} showed that for W7, the parameter q was equal to 
approximately one-third. As W7 has been able to quite successfully fit
observed spectra for normal to bright SNe~Ia 
\citep{harkness91,mazzali95,mazzali01}, we have adopted a 
q value of one-third in the calculations presented below.

With values of t$_{\circ}$, derived from the least squares fit of 
Eq.~(\ref{eq:2.0}) to the UVOIR light curve during the quasi 
exponential phase, along with the adopted fiducial values for all the 
parameters in Eq.~(\ref{eq:4.0}), we can proceed to place constraints 
on the ejected mass for all of the SN~Ia in our sample.

\subsection{$\gamma$-ray escape fraction}
By comparing the UVOIR light curve to the enrgy input from the 
radioactive decays --for both cases of complete trapping of $\gamma$ rays  
and complete escape of $\gamma$ rays-- we can obtain a quantitative 
description of the $\gamma$-ray escape fraction.
An expression for the UVOIR light curve based on this prescription can 
be written as
\beal
\label{eq:5.0}
LC(t)_{{\rm obs}} = (1 - \gamma(t))LC(t)_{\tau \gg 1} + \gamma(t)LC(t)_{\tau \ll 1}
\Abst{.}
\end{align}
In this expression LC(t)$_{{\rm obs}}$ is the UVOIR light curve, 
LC(t)$_{\tau \gg 1}$ represents the 
energy input from the radioactive decays assuming complete trapping of 
$\gamma$ rays, LC(t)$_{\tau \ll 1}$ represents the case of complete 
escape of $\gamma$ rays, and $\gamma$(t) is the $\gamma$-ray escape fraction. 
Solving Eq. (5) for $\gamma$(t) we obtain the $\gamma$-ray escape fraction
\beal
\label{eq:6.0}
\gamma(t) = \frac{LC(t)_{\tau \gg 1} - LC(t)_{{\rm obs}}}{LC(t)_{\tau \gg 1} - 
LC(t)_{\tau \ll 1}}
\Abst{.}
\end{align}

\section{Results}
\label{results}
\begin{figure}
\resizebox{\hsize}{!}{\includegraphics{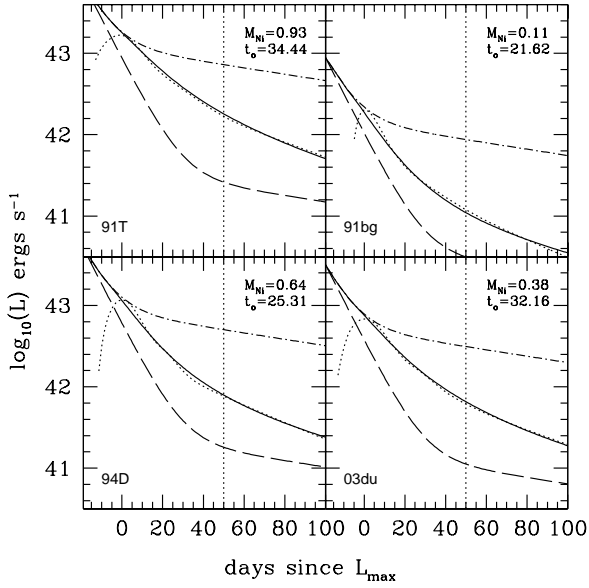}}
\caption{Fit of Eq. (2) (solid line) to the UVOIR bolometric light curve
(dotted curve) between 50 and 100 days past maximum light. 
The dashed-dotted line is 
the energy deposition of $\gamma$ rays and positrons from the
\Nif~to \Cif~to \Fif~decay, assuming complete trapping (i.e. $\tau \gg 1$). 
Dashed line is the case for complete escape of $\gamma$ rays 
(i.e. $\tau \ll 1$).
The vertical dotted line indicates the epoch ($+50$ days) when the fit begins.}
\label{fits}
\end{figure}

In Fig.~\ref{fits} we present the least squares fits of 
Eq.~(\ref{eq:2.0}) to several UVOIR light curves. 
The four events shown in Fig.~\ref{fits} are representative of the complete 
population of SN~Ia, ranging from the bright SN~1991T to the subluminous 
SN~1991bg. Also plotted are the energy deposition curves corresponding to the
\Nif~$\rightarrow$~\Cif~$\rightarrow$~\Fif~
decay chain for the cases of complete $\gamma$-ray trapping 
(dash-dotted line) and complete $\gamma$-ray escape (dashed line).
Table~1 lists the \Nif~mass calculated for each event 
through Eq.~(\ref{eq:1.0}), as well as the determined values of t$_{\circ}$.
For this sample of SNe~Ia, the \Nif~mass varies by a factor of $\sim$10, 
while t$_{\circ}$ varies by a factor of 1.6.

\begin{figure}
\resizebox{\hsize}{!}{\includegraphics{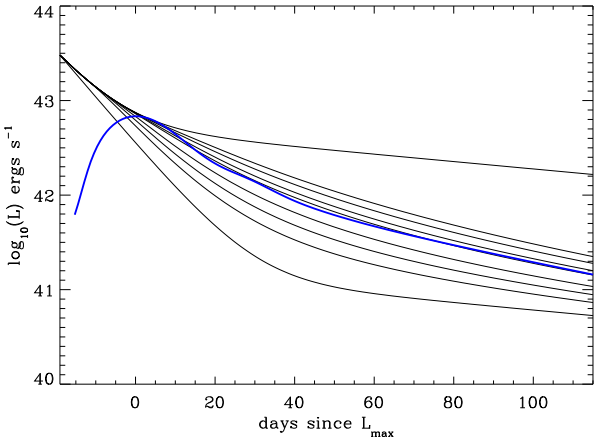}}
\caption{Plot of Eq.~(2) for a fixed \Nif~mass of 0.38~M$_{\sun}$ while 
varying t$_{\circ}$ (solid lines). Here t$_{\circ}$ ranges
(from top to bottom) $\infty$,45, 40, 35, 32.16, 25, 20, 15, and 0 days. 
The light curve corresponds to SN~2003du.}\label{ni.to}
\end{figure}

In order to give the reader a more intuitive feeling of how the
RDE deposition curve depends on the value of t$_{\circ}$, 
we present Fig.~\ref{ni.to}.
This figure contains the UVOIR light
curve of SN~2003du, along with the energy deposition curves 
for different values of t$_{\circ}$ that vary, from top to bottom:
$\infty$, 45, 40, 35, 32.16, 25, 20, 15, and 0 days.
As expected for a fixed \Nif~mass, when t$_{\circ}$ is 
increased, the energy RDE deposition function evolves more slowly
with respect to time. 
Physically this effect is associated with an increase in the diffusion time
of the photons trapped within the ejecta.

\begin{figure}
\resizebox{\hsize}{!}{\includegraphics{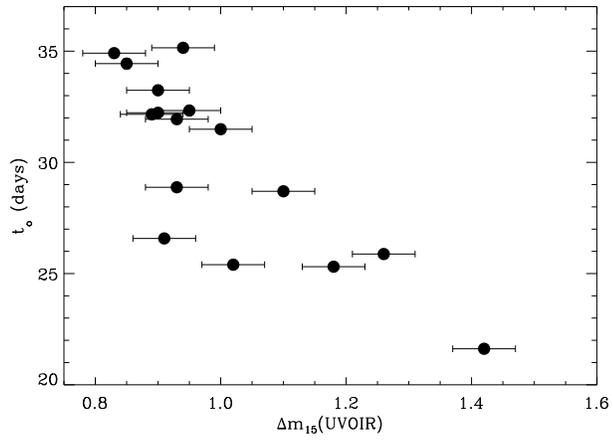}}
\caption{Fiducial time, t$_{\circ}$ plotted versus 
$\Delta$m$_{15}(\rm {UVOIR})$. 
Note the error bars associated with the values of t$_{\circ}$ are smaller 
than the size of the points.}
\label{to.m15}
\end{figure}

In Fig.~\ref{to.m15} we plot t$_{\circ}$ versus 
$\Delta$m$_{15}({\rm UVOIR})$.\footnote{By plotting 
$\Delta$m$_{15}({\rm UVOIR})$ rather 
than the \Nif~mass, we bypass the effect upon the luminosity (hence \Nif~mass)
associated with the uncertainty in the adopted distance to each event.}
Values for $\Delta$m$_{15}({\rm UVOIR})$ have been determined from the 
UVOIR light curves.
From this figure it is clear that there exists a correlation
between these two parameters. This correlation is in accord with
our expectations, as it is well established that more luminous SNe~Ia
have smaller decline rates, and thus the epoch in which their ejecta
transform to the nebular phase occurs at a later time 
\citep[see][$~$and references within for a detailed discussion of 
the physics that describes the luminosity-width relation]{pinto01}.

Armed with our values of t$_{\circ}$, we can now proceed to place constraints
on the total ejected mass. Fig.~\ref{massejecta}  
is a plot of our calculated ejected mass versus the \Nif~mass.
For the calculation of the ejected mass we have used q $= 1/3$, 
v$_{e}$ $=$ 3000~km~s$^{-1}$ and $\kappa$ $=$ 0.025~cm$^{2}$~g$^{-1}$. 

The error bars that accompany each \Nif~mass account for
uncertainties in host galaxy reddening and
the adopted distance (see Table~1). For events with a CMB
distance we have assumed 300 km s$^{-1}$ uncertainty for (random) 
peculiar velocities.

The ejected mass error bars include: (1) the uncertainty 
listed in Table~1 for each value of t$_{\circ}$, (2) a 300~km~s$^{-1}$ 
i.e. 10\% uncertainty in  v$_{e}$, (3) a 10\% uncertainty 
in $\kappa$ and (4) a 30\% uncertainty in the adopted value of q.
These `1-$\sigma$' error bars are not statistical but rather a
sensible estimation of the possible range of each parameter.

Fig.~\ref{massejecta} displays several striking features that
are worthy of comment. 
First, this figure suggests that there exists a range in the 
ejected mass of about a factor of two. Three events 
(SN~1992A, SN~1994D, and SN~2000cx) that have {\it moderate} amounts of 
M$_{ej}$ (i.e. 0.4 - 0.6 M$_{\sun}$) are of particular interest. 
These events are located nearly 3-$\sigma$ 
below the most massive events, which lie near the canonical value of 
1.4~M$_{\sun}$. In order to increase the ejected mass of these three
events to a Chandrasekhar mass,
it is necessary to reduce either q (which is highly unlikely)
or our fiducial value of $\kappa$ by a factor of two, 
or increase either the value of t$_{\circ}$ by a factor of $\sim$1.3 
or v$_{e}$ by a factor of $\sim$1.4
or more. Implementing any of these changes results in 
ejected masses for all the other `normal' SNe~Ia to be comparable 
to that of a neutron star mass.
In other words, if we change any one of the parameters in Eq~(\ref{eq:4.0})
while keeping all others constant, there will always exist a relative 
difference in the ejected mass of $\sim$2 between these three events 
shown in Fig.~\ref{massejecta}, as compared to the more massive SNe~Ia. 
Of course this is the case if
the changes are applied uniformly to the whole sample. In reality
some events may have different values 
for the parameters listed in Eq~(\ref{eq:4.0}) when compared to each other. 

The problem can, of course, be inverted to derive mean values
of q, $\kappa$, and v$_{e}$ for a fixed ejected mass.
With an ejected mass of 1.4~M$_{\sun}$ we find mean values
$<$v$_{e}>$~$= 3762$~km~s$^{-1}$,
$<$q$>$~$= 0.224$, and $<$$\kappa$$>$~$= 0.0080$~cm$^{2}$~g$^{-1}$. If the two 
subluminous events (i.e. SN~1991bg and SN~1998de) are excluded, these 
parameters change to $<$v$_{e}>$~$ $=$ 3625$~km~s$^{-1}$, $<$q$>$~$=$ $0.236$, 
and $<$$\kappa$$>$ $=$ $0.0084$~cm$^{2}$~g$^{-1}$. 

This e-folding velocity may be slightly on the high side compared
to what is predicted from successful 1-D explosion models. However, 
it is not radically different from our adopted e-folding velocity.
An explosion model with the majority of the \Nif~mixed in the outer layers,
as implied by a q~$= 0.224$ is most unlikely for the vast majority
of robust explosion models. But this could be the case for a 
Chandrasekhar mass progenitor that produces a subluminous SN~Ia.
As mentioned before, in the optically thin limit $\kappa$ ranges from
0.025 to 0.033 cm$^{2}$ g$^{-1}$. A factor of two (or even three) less of 
$\kappa$ $=$ 0.025 cm$^{2}$ g$^{-1}$ is unlikely. 
At most one could conceive of $\kappa$ to vary by $\sim50\%$. 

To calculate the amount of intermediate mass elements (IMEs)
produced during nuclear burning, we simply subtract the ejected mass from
the amount of \Nif~produced. Note this this value also includes any
stable Fe group elements and
remaining amount of unburned carbon and oxygen. These values are 
listed in Table~1.
Excluding SN~1994D we find that the IMEs range from $\sim0.11$ to
$\sim0.67~M_{\sun}$.

In Fig.~\ref{parameters} we plot the ejected mass versus t$_{\circ}$, 
while holding the other parameters of Eq.~(\ref{eq:4.0}) constant.
The solid line (case 1) corresponds to all the fiducial values used 
to determine the ejected masses in Fig.~\ref{massejecta}.
The dashed line (case 2) shows the effect of keeping q and $\kappa$ fixed
at the  fiducial values while using v$_{e}$ $=$ 3625~km~s$^{-1}$. 
For the dash-dot-dot line (case 3) we used v$_{e}$ $=$ 3625~km~s$^{-1}$,
$\kappa$ $=$ 0.0084~cm$^{2}$~g$^{-1}$ and q $=$ 0.5. Finally the dash-dot
line (case 4) corresponds to v$_{e}$ $=$ 3625~km~s$^{-1}$, 
q $=$ 1/3, and $\kappa$ $=$ 0.0084~cm$^{2}$~g$^{-1}$.
This figure illustrates the strong dependencies of the ejected masses.
Masses much above the Chandrasekhar mass are achieved for only 
extreme cases.
Both case 1 and case 2 provide ejected masses at or near the Chandrasekhar mass for 
events with large values of t$_{\circ}$ and substantially less for those
events with values of t$_{\circ}$ $\approx$ 22-26 days.

Another interesting feature displayed in Fig.~\ref{massejecta} is that there 
appears to be little or no correlation between the ejected mass 
and the amount of \Nif. This is not entirely 
unexpected because as even with the presumption that
all SNe~Ia originate from a Chandrasekhar-size white dwarf, there still
exists a range of ten or more in amount of \Nif~produced. Nevertheless
this is additional evidence which suggests that there is a 
significant variation in the burning of SNe~Ia.

\begin{figure}
\resizebox{\hsize}{!}{\includegraphics{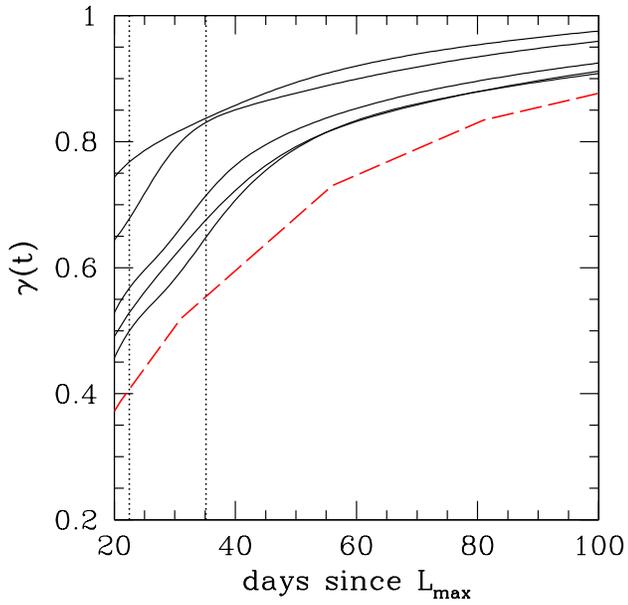}}
\caption{$\gamma$-ray escape fraction as a function of time since 
maximum light for five SN~Ia in our sample. These include (from top
to bottom) SN~1991bg, SN~1994D, SN~2003du, SN~1991T, and SN~1999dq.
The red dash line corresponds to W7. The vertical dot lines are the
minimum and maximum values of t$_{\circ}$.  
For W7 we have assumed a rise time to bolometric maximum of 19 days.}
\label{gam.esc1}
\end{figure}

We now turn our attention to the issue of the $\gamma$-ray escape fraction.
In Fig.~\ref{gam.esc1} we present the $\gamma$-ray escape fraction 
as a function of time (determined from Eq.~(\ref{eq:6.0})) for five of 
the SNe~Ia in our sample. As the
ejecta of the supernova expands there is an increase in the 
$\gamma$-ray escape fraction. This can be attributed to the decrease in 
the column density, which is accompanied with the expansion of the ejecta.
Most of the curves in this figure are accompanied by
a `bump' between 20 and 40 days past maximum light. 
Also included in Fig.~\ref{gam.esc1} is the 
$\gamma$-ray escape fraction calculated from W7.
The agreement between W7 
and our calculated $\gamma$-ray escape fraction curves 
is encouraging, considering that we are not adjusting any 
parameters.\footnote{Note that we have assumed a rise time to bolometric 
maximum of 19 days.} For the first three weeks after 
maximum light the $\gamma$-ray escape fraction 
from the UVOIR light curves is unreliable, as it is based on the assumption
of $\tau \ll 1$, which clearly is not the case for t $<$ t$_{\circ}$. 

\begin{figure}
\resizebox{\hsize}{!}{\includegraphics{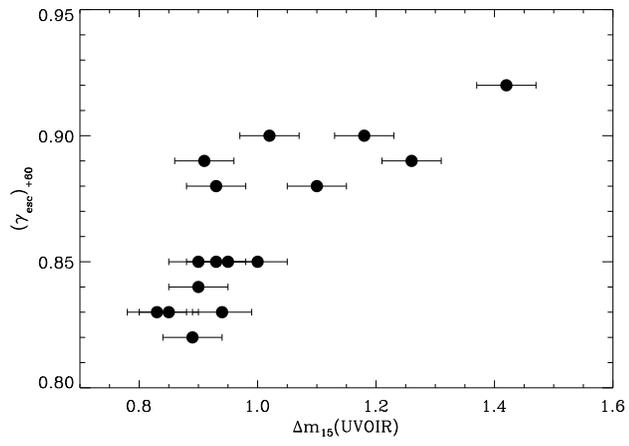}}
\caption{$\gamma$-ray escape fraction at sixty days past maximum light 
vs. $\Delta$m$_{15}({\rm UVOIR})$.}
\label{gam.esc2}
\end{figure}

From Fig.~\ref{gam.esc1} it is clear that the $\gamma$-ray escape fraction 
evolves faster in time for less luminous events. This is confirmed by
Fig.~\ref{gam.esc2} where we plot the $\gamma$-ray escape fraction at 
sixty days past bolometric maximum light versus 
$\Delta$m$_{15}({\rm UVOIR})$. At this epoch  
$\sim10\%$ more $\gamma$ rays escape in the least luminous SNe~Ia 
than in the brightest events. However, between 20 and
40 days past maximum light the differences are even more pronounced. This is 
expected, as at these epochs the UVOIR light curve has not yet reached 
its linear decline. In addition, the morphology of the secondary maximum
can vary radically from SN to SN \citep{suntzeff03,stritzinger05}.
This may then have a significant effect on the evolution of the 
$\gamma$-ray escape fraction during these phases.

\section{Discussion} 

With the stipulation that the UVOIR light curve reasonably traces
the true bolometric flux, from within the period of soon after explosion to 
one hundred days past maximum light, we have been able to derive 
constraints on several important global parameters
that relate directly to the progenitor systems of SNe~Ia.
The appeal of our approach is that, with
relative ease and simple assumptions, we have used existing data to gain 
a deeper understanding on the origins of SNe~Ia as well as provide sorely 
needed constraints on current models. 

As previously mentioned, it is commonly believed that SNe~Ia are the
result of the thermonuclear
disruption of a C-O white dwarf.
The premise that thermonuclear combustion occurs at the
Chandrasekhar limit was invoked to address the issue of homogeneity.
However, today it is well established that SNe~Ia are not true
standard candles as once thought in the past \citep[e.g.][]{leibundgut04}.\footnote{
Recently \citet{krisciunas04} have presented evidence that SNe~Ia appear to
be nearly `standard candles' in the near-infrared.}
Therefore we now must carefully scrutinize the data at hand in order 
to find plausible explanations that can account for the radical differences 
observed between different SNe~Ia.

We first showed that the \Nif~mass (hence luminosity) ranges 
from $\sim0.1$ to $\sim1.0 M_{\sun}$. 
This confirms results previously attained by several similar and independent
methods \citep{bowers97,cappellaro97,contardo00,strolger02,suntzeff03}.
But this result is quite disheartening from earlier 
assertions that SNe~Ia are standard candles. If all
SNe~Ia do indeed originate from a Chandrasekhar mass, an 
immediate question then is: What physical mechanism(s) can explain this 
range in luminosity? 

There has been considerable effort on the part of 
modelers to address this question. Yet they have been met with little
success to identify what parameter(s) can be tuned in order to account 
for a factor of ten in \Nif~mass. Obvious candidates that may affect
the production of \Nif~are the initial parameters prior to 
explosion, e.g. metallicity, central density and ignition mechanism(s).
Recently, \citet{ropke04} have shown that the C-O ratio
has essentially no effect on the amount of \Nif~produced from burning to NSE. 
If prior to explosion there is a significant amount of alpha elements within
the white dwarf, one may reasonably expect the production of more 
stable isotopes, thus reducing the amount of \Nif~synthesized 
\citep{brachwitz00}. Moreover it has been shown that changes in the central 
density of the white dwarf do influence the robustness of the explosion. 
Nevertheless, it is unrealistic that any one of these parameters, or 
even a combination of the three, can account for a factor of ten
range in the \Nif~mass. In reality these parameters affect the
production of \Nif~by no more than $\sim20\%$.

More likely to influence the amount of \Nif~synthesized is the explosion
mechanism itself \citep[see][$~$and references within for a more detailed
discussion]{stritzinger05a}. Currently the explosion mechanism and 
the subsequent evolution of the burning front is in open
debate, and varies from a subsonic deflagration to a supersonic
delayed detonation. Today the best Chandrasekhar mass models 
predict \Nif~masses that range between $\sim0.40$ to 
$0.60~$M$_{\sun}$. Due to computational 
limitations the state-of-the-art 3-D deflagration models 
\citep{reinecke02a,reinecke02b} do not produce copious amounts
of \Nif~\citep{travaglio04}, and have appreciable amounts of
unburned carbon and oxygen left over in the inner ashes \citep{kozma05}. 
The delayed detonation models
\citep{khokhlov91,woosley90,woosley94a,hoflich96}, on the other
hand, can account for some of the more luminous events, but this class 
of models requires an additional free parameter. This parameter is essential
to force the transition of the flame propagation from a deflagration to a 
detonation, and is physically not understood (however, see
\citealt{gamezo04} and \citealt{golombek05}). 
The fact that there does not exist a single class of 
Chandrasekhar mass models that can account for the complete
population of SNe~Ia is quite dissatisfying and should be seriously
addressed by theorists, if we are to insist that a Chandrasekhar-size 
white dwarf accounts for the  progenitor system of {\it all} SNe~Ia.

Under the main assumption that at times greater than fifty
days past maximum light the energy deposition in the 
ejecta of a SN~Ia is solely due to the \Cif$\rightarrow$\Fif~
decay chain, and thus the optical depth has a t$^{-2}$ dependence,
we can estimate (from the UVOIR light curve) the epoch 
when the photosphere transforms from being optically thick to 
optically thin. With this knowledge we can then use the parameterized 
SN~Ia model of \citet{jeffery99} to place constraints on the ejected mass. 

\begin{figure}
\resizebox{\hsize}{!}{\includegraphics{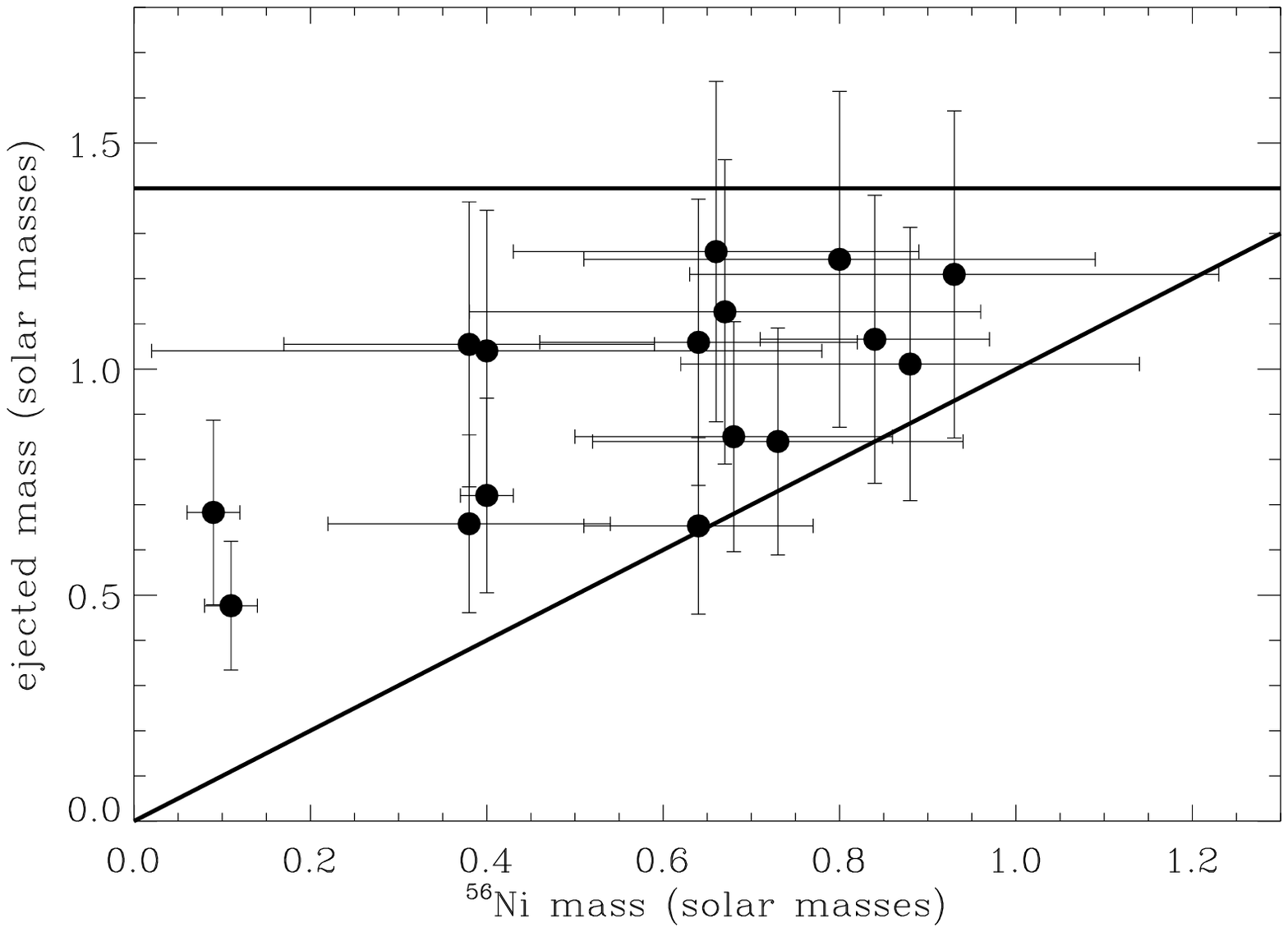}}
\caption{Ejected mass plotted vs. \Nif~mass for 16 SNe~Ia. Units
are in solar mass. See text for comments concerning the error bars.}
Solid horizontal line indicates the Chandrasekhar mass.
Slanted line has a slope of 1.
\label{massejecta}
\end{figure}

The results presented in Fig.~\ref{massejecta} provide us with  
evidence that not {\it all} SNe~Ia originate from a Chandrasekhar-size
white dwarf or other very severe differences in the explosions like the
distribution of \Nif~or kinetic energies (expansion velocities) exist. 
This would then immediately imply that some sort of
sub-Chandrasekhar mass model is responsible for at least some SNe~Ia.
If true, this would be a radical change in thinking from 
the currently favored paradigm for the progenitor systems of SNe~Ia. 
However, the suggestion that a sub-Chandrasekhar mass model
may be a viable candidate for the progenitors of some SNe~Ia is certainly 
not a new concept.
Similar to Chandrasekhar mass models, previous attempts to simulate these
systems have been plagued with their own problems. We refer the reader to
\citet{hillebrandt00} and \citet{livio00} for detailed reviews 
concerning this class of progenitor system; we briefly summarize them here.


Previous attempts to model sub-Chandrasekhar explosions
\citep{woosley94b,livne95,hoflich96}
have met with some success in reproducing the observed light curves.
However, these models typically predict a high-velocity layer 
of \Nif~and helium above the IMEs, which is not 
observed in any spectra. 
It must be noted that relatively little effort has been made to 
conduct detailed 3-D simulations of {\em sub}-Chandrasekhar mass models
(but, see \citealt{garcia99,benz97}).
With more detailed modeling, this progenitor channel may provide an 
attractive alternative to the Chandrasekhar mass model.
We also note that one appealing 
advantage offered by this model is the ability to obtain 
the progenitor statistics predicted by population synthesis calculations
\citep[see][$~$and references within]{livio00}.

Previously, \citet{cappellaro97} employed a technique that used 
observations of SNe~Ia to determine both the \Nif~mass and the ejected 
mass.
In their method they modeled the $V$-band light curves of a small sample
of SNe~Ia using a simple Monte Carlo code. 
We find that our overall results are analogous to what they determined
for both the range in the \Nif~mass and the ejected mass.
Contrary to their work we employed a different manner to 
determine these parameters and used UVOIR light curves 
rather than  V-band light curves. By using the UVOIR light curve instead 
of the $V$-band light curve, we circumvented the crude assumption that 
the latter is a close surrogate to the former during post maximum times. 
Indeed, a comparison between our UVOIR light curves to the 
$V$-band light curves indicates that by fifty days past maximum light,
the bolometric correction (m$_{{\rm bol}}$- m${_{v}}$) is $\sim0.2$ mag or 
more. At later times this difference is amplified, as the
near infrared passbands provide an increasing contribution to the
bolometric flux \citep{sollerman04}.

Although we find that our conclusions are in line with to those
presented in \citeauthor{cappellaro97}, there are subtle differences 
between the four events that coincide in both studies. 
The numbers we provide below for our results were obtained using 
Eq.~(\ref{eq:4.0}) and the fiducial values quoted previously.
Also, note that there are slight differences (no larger than
$\mu = 0.20$) in the distances used between our work and that of 
\citeauthor{cappellaro97}

For SN~1991bg, \citeauthor{cappellaro97} found a \Nif~mass 
M$_{{\rm Ni}}$ $=~0.1$~\s~and an ejected mass M$_{{\rm ej}}$ 
$= 0.7$~\s.
This is quite comparable to our findings of M$_{{\rm Ni}} = 0.11$~\s~and
M$_{{\rm ej}} = 0.48\pm0.14$~\s. 
Furthermore, our
M$_{{\rm Ni}}$/M$_{{\rm ej}}$ ratio of $0.23$  
is larger compared to their $0.14$. We found for SN~1992A,  
M$_{{\rm Ni}} = 0.40$~\s~and M$_{{\rm ej}} = 
0.72\pm0.27$\s, as compared to their M$_{{\rm Ni}} = 0.4$~\s~and 
M$_{{\rm ej}} = 1.0$~\s. This then gives us a 
M$_{{\rm Ni}}$/M$_{{\rm ej}}$ ratio of $0.56$ compared to their
$0.40$.

We find that our results with respect to the next two SNe~Ia differ
more than the first two stated events.
For SN~1994D we calculated M$_{{\rm Ni}} = 0.64$~\s~and M$_{{\rm ej}} = 
0.65\pm0.25$~\s, compared to their values of M$_{{\rm Ni}} = 0.8$~\s~and
M$_{{\rm ej}} = 1.4$~\s. Thus we obtain a larger difference in 
our M$_{{\rm Ni}}$/M$_{{\rm ej}}$ ratio of 0.98 compared to their 0.57.
The fact that we have calculated a \Nif~mass that is equal to 
the ejected mass is questionable. To determine the \Nif~mass of
SN~19994D we used a new SBF distance \citep{ajhar01} rather than the
SBF distance \citep{tonry97} used by \citet{contardo00} 
who determined a \Nif~mass of 0.40 M$_{\sun}$.
Using the distance modulus adopted by \citeauthor{cappellaro97}
we obtain a \Nif~mass of 0.67 M${\sun}$. Recently
\citet{feldmeier05} have calculated a planetary nebulae distance to 
the host galaxy of SN~1994D. In their study they have determined the distance
modulus $\mu$ $= 30.66$. This is comparable to the \citet{tonry97}
distances modulus $\mu$ $= 30.68$. Using the planetary nebulae distance
the \Nif~mass would be reduced to $\sim0.40 M_{\sun}$.
Nonetheless the \Nif~mass determined by us and \citeauthor{contardo00} 
is less than the 0.8 M$_{\sun}$ calculated by \citeauthor{cappellaro97}
with their method. The discrepancies between these values of 
the \Nif~mass underscores the effect of the uncertainty in the distances.

Finally, for SN~1991T \citeauthor{cappellaro97} assumed 
M$_{{\rm ej}}$ $=$ M$_{{\rm Ni}}$ where M$_{{\rm Ni}}$ $=$ 1.1~\s.
We, on the other hand, found M$_{{\rm Ni}} = 0.93$~\s~and 
M$_{{\rm ej}}$ $=$ $1.21\pm0.36$~\s. In summary we find the results
presented by \citeauthor{cappellaro97} to be in fair agreement with our 
calculations, although some discrepancies do exist.    

We have presented an investigation of the bolometric behavior of sixteen 
SNe~Ia. 
In particular we have provided important constraints on the progenitor
system(s) of these stellar explosions. Our 
results suggest that some progenitor system(s) of SN~Ia
may emanate from the thermonuclear explosion of a 
sub Chandrasekhar-size white dwarf. This result may be difficult to reconcile
with the current paradigm of the progenitor system of SNe~Ia, i.e. a
Chandrasekhar-size white dwarf. Moreover, our results suggest
that the amount of \Nif~produced during the explosion is most likely 
not dependent on the mass of the progenitor, but more likely
on the manner in which nuclear burning is initiated and the
subsequent dynamics of the flame propagation through the white dwarf. 
The range in synthesized \Nif~possibly indicates that there are two
different explosion mechanisms.
Further modeling of the explosion mechanism is required in order to 
investigate how different initial conditions can affect the 
observed range in luminosity.

\begin{figure}
\resizebox{\hsize}{!}{\includegraphics{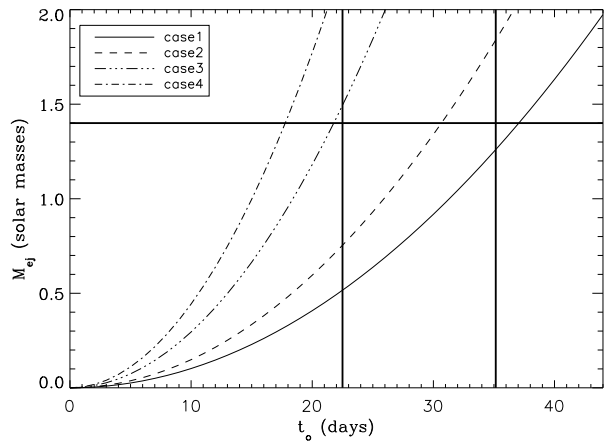}}
\caption{Ejected mass plotted vs. t$_{\circ}$ for fixed values
of the parameters in Eq~(\ref{eq:4.0}). See Sect.~\ref{results} for 
a complete description of each curve. Solid vertical lines indicates
the minimum and maximum values for t$_{\circ}$ in our sample, and the 
solid horizontal lines indicates the Chandrasekhar mass.}
\label{parameters}
\end{figure}

In Fig.~\ref{massejecta} we see --in contrast to current thinking--
that the mean ejected mass of many explosions is on the low side. 
A valid concern is that the parameters used to determine the
ejected masses may not exactly represent those of a real
SN~Ia explosion. One parameter that could be in 
error, and does have a significant effect on our estimates of the 
ejected masses, is the adopted value of the e-folding
velocity (see Fig~\ref{parameters}). If we assume slightly larger 
values of v$_{e}$, the mean ejected mass for our sample would be in better
agreement with 1.4~M${\sun}$. How this parameter differs
in 3-D simulations compared to 1-D simulations is not yet clear. 

In addition, the simple assumption that any one of the parameters in 
Eq.~(\ref{eq:4.0}) 
is unique for all events is probably incorrect. This may have a significant 
effect on the determined ejected mass for each event. However, this does 
not necessary imply that we would obtain larger ejected masses. It would be 
helpful if the theorists in the future provided values of 
v$_{e}$ and q from their simulations.

An acceptable argument concerning
the results presented in this work is the 
validity of the model used to determine the ejected 
mass. There may be several assumptions built into the
parameterized model of \citet{jeffery99} 
which may be too na\"ive, and therefore the model 
may not adequately account for various complicated
physical processes that occur within the progenitor
of a SN~Ia. However, there currently exists no other
method to use observed photometry 
to place constraints on such a parameter. 

In principle we would like to compare the UVOIR light curves 
to detailed NLTE modeled light curves.
Unfortunately there has been little success in 
such an endeavor, owing to the complications in 
performing such time-dependent calculations as well as the limits 
of atomic line data, however see \cite{kozma05}. The next step 
will be to fit UVOIR light curves to a grid of model light curves produced
from 3-D radiative transfer calculations, and then place further
constraints on the progenitor systems of SNe~Ia.

\begin{acknowledgements}
M.S. acknowledges the International Max-Planck Research School
on Astrophysics for a graduate fellowship. M.S. is grateful for helpful 
discussions with Sergei Blinnikov, Wolfgang Hillebrandt, Gert H\"utsi, 
Paolo Mazzali, Brian Schmidt, and the SNe~Ia group at the MPA. 
This research has made use of the NASA/IPAC 
Extragalactic Database (NED), which is operated by the Jet Propulsion 
Laboratory, California Institute of Technology, under contract with the 
National Aeronautics and Space Administration.
\end{acknowledgements}

\clearpage
\begin{deluxetable}{l l l c c c c c c c c c}
\rotate
\tablecolumns{12}
\tablenum{1}
\tablewidth{0pc}
\tablecaption{Well-observed SNe~Ia}
\label{data.tab}
\tablehead{
\colhead{SN} &
\colhead{Filters} &
\colhead{Ref.} &
\colhead{E(B-V)$_{{\rm gal}}^{a}$} &
\colhead{E(B-V)$_{{\rm host}}$} &
\colhead{$\mu$} &
\colhead{vel} &
\colhead{Ref.$^{b}$} &
\colhead{M$_{{\rm Ni}}$} &
\colhead{t$_{0}$} &
\colhead{M$_{{\rm ej}}$} &
\colhead{M$_{{\rm IME}}$}\\
\colhead{} &
\colhead{} &
\colhead{} &
\colhead{} &
\colhead{} &
\colhead{} &
\colhead{(km s$^{-1}$)} &
\colhead{} &
\colhead{(M$_{\sun}$)} &
\colhead{(days)} &
\colhead{(M$_{\sun}$)} &
\colhead{(M$_{\sun}$)}
} 
\startdata
SN1989B  & $UBVRI$ & 1         & 0.032 & 0.340(0.04) & 30.22(0.12)  & 797  & 21 & 0.64(0.18) & 32.23(0.12) & 1.06(0.32) & 0.42 \\ 
SN1991T  & $UBVRI$ & 2         & 0.022 & 0.140(0.05) & 30.74(0.12)  & 1012 & 22 & 0.93(0.30) & 34.44(0.23) & 1.21(0.36) & 0.28 \\
SN1991bg & $BVRI$  & 3, 4, 5   & 0.040 & 0.030(0.05) & 31.32(0.11)  & 1322 & 23 & 0.11(0.03) & 21.62(0.11) & 0.48(0.14) & 0.37 \\
SN1992A  & $UBVRI$ & 6         & 0.017 & 0.000(0.02) & 31.35(0.07)  & 1341 & 24 & 0.40(0.03) & 26.58(0.10) & 0.72(0.22) & 0.32 \\
SN1994D  & $UBVRI$ & 7, 8, 9   & 0.022 & 0.000(0.02) & 31.08(0.20)  & 1184 & 25 & 0.64(0.13) & 25.31(0.11) & 0.65(0.20) & 0.01 \\
SN1994ae & $BVRI$  & 10        & 0.031 & 0.120(0.03) & 32.29(0.06)  & 2067 & 13 & 0.84(0.13) & 32.33(0.13) & 1.07(0.32) & 0.23 \\
SN1995D  & $BVRI$  & 10, 11    & 0.058 & 0.040(0.02) & 32.50(0.26)  & 2272 & CMB& 0.66(0.23) & 35.15(0.12) & 1.26(0.38) & 0.60 \\
SN1995E  & $BVRI$  & 10        & 0.027 & 0.740(0.03) & 33.42(0.18)  & 3478 & CMB& 0.88(0.26) & 31.49(0.11) & 1.01(0.30) & 0.13 \\ 
SN1996X  & $UBVRI$ & 10, 12    & 0.069 & 0.010(0.02) & 32.40(0.20)  & 2174 & 12 & 0.73(0.21) & 28.70(0.10) & 0.84(0.25) & 0.11 \\
SN1998aq & $UBVRI$ & 13        & 0.014 & 0.002(0.05) & 31.66(0.09)  & 1547 & 13 & 0.68(0.18) & 28.88(0.17) & 0.85(0.25) & 0.17 \\
SN1998de & $BVRI$  & 14        & 0.060 & 0.000(0.05) & 34.05(0.14)  & 4653 & CMB& 0.09(0.03) & 27.80(0.10) & 0.68(0.20) & 0.59 \\
SN1999ac & $UBVRI$ & 15        & 0.046 & 0.120(0.05) & 33.06(0.21)  & 2949 & CMB& 0.67(0.29) & 33.24(0.11) & 1.13(0.34) & 0.46 \\
SN1999dq & $UBVRI$ & 16        & 0.024 & 0.139(0.05) & 33.74(0.16)  & 4029 & CMB& 0.80(0.29) & 34.91(0.14) & 1.24(0.37) & 0.44 \\
SN2000cx & $UBVRI$ &16, 17, 18 & 0.082 & 0.000(0.05) & 31.90(0.20)  & 1727 & 23&  0.38(0.16) & 25.40(0.09) & 0.66(0.20) & 0.28 \\
SN2001el & $UBVRI$ & 19        & 0.014 & 0.206(0.05) & 30.83(0.54)  & 1053 & CMB& 0.40(0.38) & 31.94(0.12) & 1.04(0.31) & 0.64 \\
SN2003du & $UBVRI$ & 20        & 0.010 & 0.000(0.05) & 32.23(0.30)  & 2011 & CMB& 0.38(0.21) & 32.16(0.13) & 1.05(0.32) & 0.67 \\

\enddata
\tablenotetext{a} {Taken from \citet{schlegel98} dust maps.}
\tablenotetext{b} {For events without a direct distance estimate we selected Heliocentric velocities listed in NED and transformed these to the CMB reference frame. To account for peculiar velocities we assume throughout this work an error
of 300 km s$^{-1}$ for all CMB distances.}

\tablerefs{
(1)~\ \citet{wells94}, (2)~- \citet{lira98}, (3)~- \citet{filippenko92}, (4)~- \cite{leibundgut93}, (5)~- \citet{turatto96}, (6)~- \citet{suntzeff96}, (7)~- \cite{richmond95}, (8)~- \citet{meikle96}, (9)~- \citet{smith00}, (10)~- \citet{riess99a}, (11)~- \citet{sadakane96}, (12)~- \citet{salvo01}, (13)~- \citet{riess05}, (14)~- \citet{modjaz01}, (15)~- \citet{phillips03}, (16)~- \citet{jha02}, (17)~- \citet{li01}, (18)~- \citet{candia03}, (19)~- \citet{krisciunas03}, (20)~- \citet{stanishev05}, (21)~- \citet{saha99}, (22)~- \citet{saha01a}, (23)~- \citet{tonry01}, (24)~- \citet{madore99}, (25)~- \citet{ajhar01}.} 

\end{deluxetable}

\clearpage

\clearpage

\clearpage

\clearpage

\clearpage

\clearpage

\end{document}